\documentstyle[11pt]{article}

 \title{
\Large\bf \begin{center}
Physics of Gravitational Interaction:
\end{center}
 \begin{center}
Geometry of Space or Quantum Field in Space?
\end{center} }
 \author{
  \bf{Yurij Baryshev}
\\
 Astronomical Institute of the St.Petersburg University,\\
  198504 St.Petersburg, Russia,\\
 e-mail: yuba@astro.spbu.ru}

\date{~}

\begin{document}

\maketitle

\begin{abstract}
\noindent

Gravity theory is the basis of modern cosmological models.
Thirring-Feynman's tensor field approach to gravitation 
is an alternative to General Relativity (GR).
Though  Field Gravity (FG) approach is still developing subject,
it opens new
understanding of gravitational interaction,
stimulates novel experiments on the nature of gravity and gives
possibility to construct new cosmological models in Minkowski space.
According to FG, the universal gravity force
is caused by exchange of gravitons - the quanta of gravity field.
Energy of this field is well-defined and excludes the singularity.
All classical relativistic effects are the same as in GR, though
there are new effects, such as free fall of rotating
bodies, scalar gravitational radiation, surface of relativistic
compact bodies, which may be tested experimentally.
The intrinsic scalar (spin 0) part of gravity field corresponds
to "antigravity" and only together with the pure tensor
(spin 2) part gives the usual Newtonian force. Laboratory
and astrophysical experiments for testing new predictions of FG,
will be performed in near future.  In particular observations 
with bar and interferometric detectors,
like Explorer, Nautilus, LIGO and VIRGO, will check the predicted
scalar gravitational waves from supernova explosions. 

\end{abstract}

\maketitle

%%%%%%%%%%%%%%%%%%%%%%%%%%%%%%%%%%%%%%%%%%%%
%% MAINMATTER
%%%%%%%%%%%%%%%%%%%%%%%%%%%%%%%%%%%%%%%%%%%%

\section{What is gravity?}
Physical understanding of fundamental interactions - strong,
weak, and electro-magnetic, is based on the quantum fields.
Then why is the most evident
interaction in the world - gravity - not in the
above list?  In fact, this question
concerns the nature of gravitational interaction.
Is gravity a manifestation of material field,
whose exitations - gravitons, are responsible for the fall of
Newton's apple? Or, is it a reflection of the geometry
of space, whose curvature gives the apple the natural
state of free fall? In other words: is gravity a kind of
matter in space, or is it the curved space itself?

The generally accepted answer is that gravity 
is interpreted in the framework of a 
geometric theory -- General Relativity (GR),
which has a successful history of experiments, but only in
weak gravity conditions. Also, GR is not a quantum theory.

In this report I discuss another possible answer that gravity
may be described as a material relativistic field in Minkowski
space, which is called Field Gravity (FG) approach. 
I shall argue that field
understanding of gravity opens
new possibilities for experimental testing of the nature of
gravity. Also new types of cosmological models in Minkowski space
are possible.

\section{Thirring-Feynman field approach to gravitation}

In 1960's Thirring (1961) [13] and 
Feynman (1971, 1995) [6, 7] made an attempt
to describe gravity as a relativistic tensor field in Minkowski space,
using Lagrangian formalism of the field theory.
In his \emph{Lectures on Gravitation} Richard Feynman  discussed
a standard quantum field description of gravity "just as the next
physical interaction". He emphasized that "the geometrical
interpretation is not really necessary or essential to physics"
([6], p.110, Lecture 8). Hence Feynman's field gravity approach is
a natural starting point for understanding the physics of gravity phenomena
similarly as other fundamental forces.

According to FG approach, the gravity force between the proverbial
 Newtonian apple and the Earth
 is caused by the exchange of gravitons. Gravitons
 (real and virtual) are mediators of the
gravitational interaction and actually represent the quanta
of the relativistic tensor field $\psi^{ik}$ in Minkowski space
$\eta^{ik}$.

Note that the construction of FG is not completed and many
important questions are still open. Only weak field approximation
is studied in detail but this is enough to demonstrate
feasibility of the FG approach and to find new predictions
which may distinct FG and GR [2].

FG approach naturally brings about the solution of the
long standing energy problem of General Relativity. It is
well known (see e.g. Landau \& Lifshitz: \emph{The classical theory
of fields}, 1971 [8], p.304 ) that in GR there is no satisfactory concept
of energy-momentum tensor (EMT) of gravity (pseudotensor is not
a tensor). However in the FG
the Minkowski space implies the
invariance under the Poincar\'{e} group transformation and hence
the usual definition of the gravity
field EMT, which follows from Noether's theorem.
 This well-defined EMT allows one consistently to consider
energy quanta of gravity field, which transmit the gravity force.

The first step in the construction of the FG is to choose
the Lagrangians for the free field and for the interaction in the low
energy regime. This was done for the case of 
weak field and the result is relativistic
tensor field gravity theory in Minkowski space, first suggested by
Poincar\'{e} and Birkhoff, and futher considered by Thirring and Feynman. 
Discussion of main 
equations and predictions of the Field Gravity approach are presented 
in [2].  Here I
highlight the most interesting points which FG approach
uncovers for the physics of gravity.

 It should be noted that in the literature
one may often find incorrect claims about the field approach to gravity,
such that it is impossible to construct a consistent field gravity theory
 or that
FG and GR are completely equivalent
(see e.g. Misner, Thorn \& Wheeler 1973 [9] sections 7 and 17). 
In his "Lectures on Gravitation",
where Feynman discussed the initial principles of the field approach,
after consideration of the weak field he simply jumped to exact 
equations of geometrical GR. Then Deser (1970) claimed that he
derived exact GR equations starting from weak case of the field
approach. However in his iteration procedure he used an expression
for EMT of the gravity field which did not satisfy to the basic
conditions of bosonic zero-mass particle EMT.

The principal difference between GR and FG was demonstrated
by Baryshev (1996) [2] and Straumann (2000) [12] -- the main point
is that in curved geometry there is no conserved energy-momentum tensor
of gravity field. Also in FG there is no complex 
topology of space as it is in GR.

\section{Surprises of the low energy regime}

Let us consider the case of weak gravity field. 
A widespread mistake
is the statement that the symmetric tensor field corresponds to
particles with spin 2 only.
Indeed, the multicomponent structure of the tensor potential
is one of the most important things for quantum field theory.
In the case of the symmetric second rank tensor field $\psi^{ik}$
there are 10 independent components which represent particles with spins
two, one, and zero (two particles with spin 0), participating in virtual
quantum processes:

\[\{\psi^{ik}\} = \{2\}\oplus\{1\}\oplus\{0\}\oplus\{0\}~~.\]

Because the field equations are gauge invariant
under the transformation $\psi^{ik} \longrightarrow \psi^{ik} +
\lambda^{i,k} + \lambda^{k,i}$
 one may
use 4 additional functions $\lambda^i$ to delete 4 components corresponding
to spin 1 and the first spin 0, so leaving only the spin 2 and 
the second spin 0
parts of the tensor potential:

\[\{\psi^{ik}\} = \{2\}\oplus\{0\}~~.\]

This means that tensor FG theory is actually
a scalar-tensor theory, where the scalar part of the field
is simply the trace $\psi =\eta_{ik} \psi^{ik}$ and it
corresponds to the scalar part of the source $T=\eta_{ik}T^{ik}$.
Hence in the case of free field the potential
is the sum of two independent parts: a pure tensor wave with spin 2 and
a scalar wave with spin 0.

The scalar $\psi$ is an intrinsic part of the gravitational tensor
potential $\psi^{ik}$ and does not relate to extra scalar fields
usually introduced in the Jordan-Fierz-Brans-Dicke theories.
So all constraints on the extra scalar field $\varphi$, 
do not restrict the scalar part $\psi$ of the tensor field $\psi^{ik}$.

The most intriguing consequence of the Field Gravity is that
the scalar part (spin 0) corresponds to a repulsive force,
while the pure tensor part (spin 2) corresponds to attraction.
Hence the usual Newtonian force is actually the sum of
two parts where the attractive force is three times the repulsive
force [2, 3]:

\[\vec{F_N}=(\vec{F_{(2)}}+\vec{F_{(0)}})=
-\frac{3}{2}m\vec{\nabla}\varphi_N + \frac{1}{2}m \vec{\nabla}\varphi_N
= -m\vec{\nabla}\varphi_N~~,\]
which directly follows from the equation of motion
for test particles. This understanding of the Newtonian
potential opens new ways for experimental studies
of the scalar "antigravity" even in the weak field laboratory conditions.

The field equation for the scalar part $\psi$ is the usual wave
equation:

\[(\bigtriangleup - \frac{1}{c^2}\frac{\partial^2}{\partial t^2})
\psi(\vec{r},t) = -\frac{8\pi G}{c^2}T(\vec{r},t)~~.\]
which describes the generation of scalar gravitational waves
by the trace of the EMT of the source. Scalar gravitational waves are
longitudinal and are generated by spherical pulsations
of the source [1]. This essentially influences the supernova explosion 
phenomenon
and the predictions for expected signals
in gravitational wave antennas [4, 10, 11].

\section{Universality of gravity force}

The basic principles of the Field Gravity are the same as
for other relativistic quantum fields. 
These include the Minkowski space, the
quantum uncertainty principle and the many-path approach.

The equivalence principle of GR cannot be a basis of the
Field Gravity, because it eliminates gravity force and accepts
the equivalence between inertial motion and accelerated motion
under gravity. E.g. the equivalence principle creates such a puzzle
as the radiating electric charge resting in the Earth's gravity field
on a laboratory
table, just due to the equivalence
of this frame to a constant acceleration of
the table.

The concept of inertial frame is preserved in Field Gravity
and there is no equivalence between inertial and accelerated
motions. Instead of the principle of equivalence, FG is based on
the principle of universality of gravitational interaction, first
formulated by Marcos Moshinsky in 1950 as the universal form of
the interaction Lagrangian:

\[\Lambda_{int}=-\frac{1}{c^2}\psi_{ik}T^{ik}~~.\]

As a consequence of this principle, inertial and gravitational
masses of a usual test body are simply equal to its rest mass.
The charge on the table does not radiate, and all classical
relativistic gravity effects have the same values in FG and GR.

The equation of motion of a test particle in a static weak field
in Post-Newtonian approximation is

\[\frac{d\vec{v}}{dt}=-(1+\frac{v^2}{c^2} + 4\frac{\varphi_N}{c^2})
\vec{\nabla}\varphi_N + 4\frac{\vec{v}}{c}(\frac{\vec{v}}{c}
\vec{\nabla}\varphi_N)~~.\]

From this equation immediately follows a possible generalization
of the old Galileo-Stevinus experiment. Such 
a "Galileo-2000 experiment" would test the fact that the gravitational
 force
acting on a test particle depends on the value and direction of its
velocity. Hence rotating bodies with differently oriented
angular momenta will fall with different accelerations.
Another version of this experiment is weighing rotating bodies
with a balance scale, where one can measure directly the difference
in gravity forces.  For astronomical binary systems this effect will
appear as a periodical modulation of the orbit of a rapidly rotating
body (Baryshev 2002 [3]).

\section{The absence of singularities in Field Gravity}

The energy density of the gravitational field in FG theory for the
case of a static weak field is

\[\varepsilon_g = T^{00}_g = \frac{1}{8\pi G}(\vec{\nabla}\varphi_N)^2~~.\]
It is positive, localizable, and does not depend on a choice of the
coordinate system.

A very general energy argument leads in FG to exclusion of singularities.
 Indeed, the total energy of the gravity field
existing around a body, should be less than the rest
mass energy of the body:

\[E_g < Mc^2  \Rightarrow  R_o > GM/2c^2~~.\]

Thus black holes and singularities are excluded by the energy of
 gravity field.
This argument is a precise analogue to that of the classical
radius of electron $R_e>e^2/m_ec^2$, following from the requirement
that the field energy $E_e$ should be less than the electron's
rest-mass energy.

Instead of black holes in Field Gravity there are compact relativistic
objects having radiuses close to gravitational radius $R_m=Gm/c^2$. 

\section{Astrophysical tests of Field Gravity}

All classical relativistic gravity effects in the Solar System
and binary pulsars up to now do not probe the genuine difference
between FG and GR. Though, also some differing effects
exist even in the weak gravity field, such as the periodical
modulation of the orbit of a rapidly rotating body 
(Baryshev 2002 [3]).

In the case of strong gravity the predictions of FG and GR diverge
dramatically. In FG there is no black holes and singularities,
and no such limit as the Oppenheimer-Volkoff mass. This means that
compact massive objects in binary star systems and active
galactic nuclea are good tests for the FG theory 
(Baryshev 1996 [4]).

Scalar gravitational radiation is predicted by the FG for the
spherical pulsations of exploding cores of massive supernovae.
This prediction will be tested in a few years by the new-generation
gravitational antennas 
(Baryshev 1995 [1];
Baryshev \& Paturel 2001 [4]; 
Paturel \& Baryshev 2003a,b [10, 11]) .

In cosmology FG provides the possibility 
to study infinite matter distributions
in Minkowski space, without the gravity paradox, and
naturally gives the zero curvature models.
  The EMT of the interaction
plays the role of an effective cosmological $\Lambda$-term
(see Baryshev et al. 1994 [5]).
The possibility of a non-zero rest mass of the graviton may lead to
new cosmological solutions. 

\section{Conclusions}

Feynman's field approach to gravity clearly deserves more attention.
It opens new
understanding on the physics of gravitational interaction and
gives new ideas for the development of gravity experiments
in the laboratory and the cosmos.

It seems that general relativity will be included partly
in the frame of the quantum field theory as a classical limit,
similarly as geometrical optics is a limiting case of quantum
electrodynamics.  Further experiments will show which principle is
more fundamental: the uncertainty principle of quantum physics or
the principle of equivalence of general relativity.

In his third letter to Bentley, Newton wrote: "Gravity must be caused
 by an Agent acting constantly
according to certain Laws; but whether this Agent be material
or immaterial, I have left to the Consideration of my Readers."[14]
Because of the testable predictions of general relativity and field
gravity, one may
hope that the forthcoming astrophysical observations will be able
to answer Newton's question - What is the nature of gravity? \\

\begin{center}
{\bf Acknowledgments}
\end{center}

The author is grateful to the 
 Domingos Silva Teixeira
for financial support. \\

\begin{center}
{\bf REFERENCES}
\end{center}

%{\bf REFERENCES}
%\begin{thebibliography}{99}
%\newline

1. Baryshev Yu., 
On a possibility of scalar gravitational wave detection
from the binary pulsar PSR 1913+16,
Proceedings of the First Edoardo Amaldi Conference on
\emph{Gravitational wave experiments}, eds. E.Coccia,
G.Pizzella, F.Ronga,
World Scientific Publ.Co., 1995, p.251
(gr-qc/9911081).

2. Baryshev Yu.,
 Field Theory of Gravitation:
Desire and Reality, Gravitation,v.2,p.5, 1996 (gr-qc/9912003).

3. Baryshev Yu., 
Translational motion of rotating bodies and
tests of the equivalence principle.
Gravitation \& Cosmology, vol.8, 232, 2002 (gr-qc/0010056).

4. Baryshev Yu., Paturel G.
Statistics of the detection rates for tensor and scalar gravitational
waves from the local galaxy universe.
Astron. Astrophys., vol.371, 378-392, 2001.

%\bibitem[1994]{Baryshev94}
5. Baryshev, Yu., Sylos Labini F., Montuori, M., Pietronero, L.,
Facts and ideas in modern cosmology,
Vistas in Astronomy 38, 419, 1994.

6. Feynman R., \emph{Lectures on Gravitation} ,1962-63,
California Institute of Technology, 1971.

7. Feynman R., Morinigo F., Wagner W.,
 \emph{Feynman Lectures on Gravitation},
Addison-Wesley Publ. Comp., 1995.

%\bibitem[1971]{Landau}
8. Landau, L.D., Lifshitz, E.M.,\emph{ The Classical Theory of Fields},
(Pergamon Press), 1971.

9. Misner C., Thorn K., Wheeler J.,
\emph{Gravitation},
(W.H. Freeman and Company), 1973.

10. Paturel G., Baryshev Yu.,
Sidereal time analysis as a tool for the space distribution
of sources of gravitational waves, 
Astron. Astrophys., vol.398, 377, 2003. 

11. Paturel G., Baryshev Yu.,
Prediction of the sidereal time distribution of gravitational
wave events for different detectors
Astrophys. J. Lett., vol.592, L99, 2003. 

12. Straumann N.,
Reflections on gravity
2000 (astro-ph/0006423).

13. Thirring W. E.,
 An alternative approach to the
 theory of gravitation,
Ann. of Phys., v. 16,  96, 1961.

14. Newton Isaac, 
From \emph{Four Letters from Sir Isaac Newton to Doctor Bentley},
London: Printed for R. and J. Dodsley, Pall-Mall, M DCC LVI (p.26)

\end{document}